\documentclass[%
 aip,
 amsmath,amssymb,
 reprint,%
]{revtex4-1}

\usepackage{graphicx}
\usepackage{dcolumn}
\usepackage{bm}

\usepackage[utf8]{inputenc}
\usepackage[T1]{fontenc}
\usepackage{mathptmx}
\usepackage{xcolor}

\begin{document}

\preprint{AIP/123-QED}

\title{Can broken symmetry spin a particle using a linearly polarized tightly focused Gaussian beam?}


\author{Sauvik Roy}
\email{sauvikroy3388@gmail.com}
\affiliation{Department of Physical Sciences, IISER-Kolkata, Mohanpur 741246, India}

\author{Ram Nandan Kumar}
\affiliation{Department of Physical Sciences, IISER-Kolkata, Mohanpur 741246, India}

\author{Biswajit Das}
\affiliation{Department of Physical Sciences, IISER-Kolkata, Mohanpur 741246, India}

\author{Nirmalya Ghosh}
\affiliation{Department of Physical Sciences, IISER-Kolkata, Mohanpur 741246, India}

\author{Subhasish Dutta Gupta}
\affiliation{Department of Physical Sciences, IISER-Kolkata, Mohanpur 741246, India}
\affiliation{Tata Institute of Fundamental Research, Hyderabad, Telangana 500046, India}

\author{Ayan Banerjee}
\email{ayan@iiserkol.ac.in}
\affiliation{Department of Physical Sciences, IISER-Kolkata, Mohanpur 741246, India}
\date{\today}

\begin{abstract}
Conventional knowledge rules out optical torque-induced spinning/rotational motion of birefringent microparticles by a linearly polarized Gaussian beam, carrying zero net spin angular momentum. Even when tightly focused, despite having strong spin orbit interaction, the axial symmetry of the system does not allow non-zero net spin angular momentum, and hence no optical torque near the focal plane. Here, we demonstrate that the limitation can be overcome using several commonly encountered symmetry breaking modalities (eg. tilt, shift and shape of the beam) of the focusing system, thereby disrupting the symmetric separation of the right and left circular spin components for the same linearly polarized Gaussian beam. As a consequence, trapped microparticles experience a tunable optical torque and exhibit rotational motion with distinct rotational frequencies at the same input power. The particles also undergo controlled reversal of the rotation direction simply by rotating the incident plane of polarization using a half-wave plate. Despite their apparent diversity, all these methods share the same physical origin rooted in asymmetric illumination. These asymmetries govern the competition between the alignment and spin-mediated torques, with continuous rotation emerging when the spin-mediated torque dominates over the alignment torque. These results establish an experimentally accessible and minimal strategy for realizing controllable optical rotation devices exploiting spin-orbit optomechanics driven by light which need not possess intrinsic spin angular momentum.
	
\end{abstract}

\maketitle


\section{Introduction}
\begin{figure*}
\centering\includegraphics[width=0.7 \linewidth]{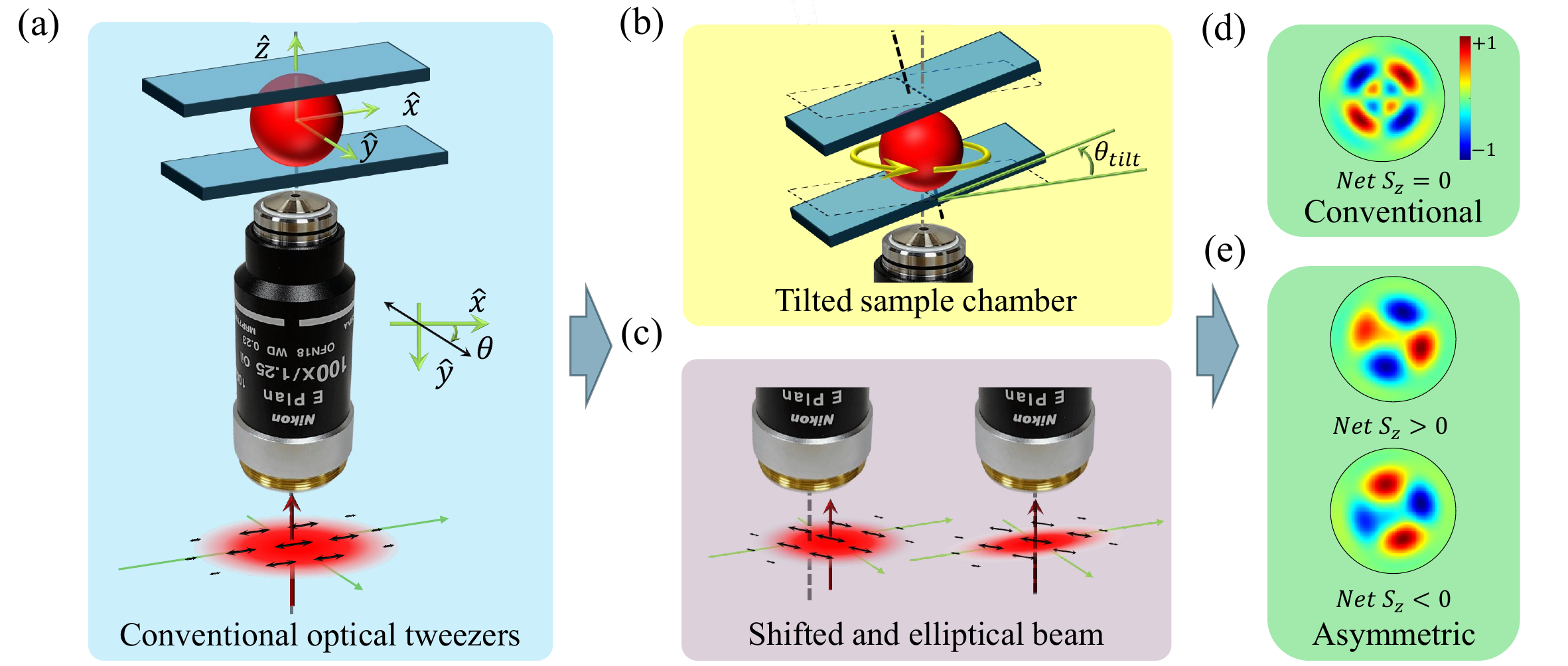}
\caption{Schematic illustration of (a) conventional optical tweezers and asymmetric optical tweezers realized through (b) a tilted sample chamber and (c) laterally shifted or elliptically shaped beam cross-sections. These symmetry-breaking configurations breaks the symmetry of the (d) longitudinal spin angular momentum (LSAM) distribution, giving rise to (e) a net positive or negative LSAM ($S_z$). The sign of the net LSAM is governed exclusively by the plane of polarization of the incident linearly polarized Gaussian beam, thereby producing positive or negative optical torque on the trapped birefringent particle and inducing counterclockwise or clockwise rotation, respectively.}
\label{fig:schematic}
\end{figure*}

Light carries energy, momentum, and angular momentum. The interplay between intrinsic spin angular momentum (SAM) \cite{bliokh2015physreports}, intrinsic and extrinsic orbital angular momentum (IOAM and EOAM) \cite{IEOAMpadgett}, and their mutual interconversions mediated by the spin–orbit interactions of light (SOI) \cite{bliokh2naturephotonics,ramnandanadvancedphotonicsnexux,ramnandanpra,sramanaanal,Bliokhfocusingscatteringimaging,sauvikapl,sauvikpra,snigdhadevjofphysicscomm} has given rise to a wealth of striking optical phenomena. These include the generation of spin-dependent optical vortices, SAM- and OAM-dependent beam trajectory shifts—known as the spin and orbital Hall effects of light \cite{bliokh2015quantum,roy2014manifestations,Bliokh:19,PhysRevLett.112.113902,Remesh2022,ramrotshe,kotlyar2023hall} -- as well as the emergence of transverse spin angular momentum (TSAM) \cite{bliokh2014extraordinary,neugebauer2015measuring,neugebauer2018magnetic,aiello2015transverse,saha2016transverse,mukherjee2016coherent,PhysRevLett.103.100401,saha2018JOP} in evanescent \cite{bliokh2014extraordinary} and tightly focused optical fields \cite{PhysRevA.87.043823,svak2018transverse,pal2020direct,saha2016transverse,saha2018transverse}. 
In the context of light-driven dynamics of matter at the meso- and nanoscale, optical forces and optical torques originating from the transfer of linear and angular momentum of light provide a unified and physically intuitive framework for describing particle motion. Within optical tweezers (Fig. \ref{fig:schematic} (a)), optical torque complements the gradient and scattering forces by enabling simultaneous control over both translational and rotational degrees of freedom, allowing particles to be positioned, oriented, and rotated with high precision \cite{nieminen2007physics,pesce2020optical,padgett2011tweezers,haldar2012self,PhysRevA.87.043823,ramnandanadvancedphotonicsnexux,sauvikapl,snigdhadevjofphysicscomm}.

Conventionally, left- or right-circularly or elliptically polarized Gaussian beams are employed for this purpose, with the mutual interconversion of the SAM and OAM leading to exotic rotational effects in mesoscopic particles~\cite{padgett2011tweezers,rotationdraghalina}. Linearly polarized light has also been used for generating spinning of particles using the effects of the optical spin-Hall effect amplified by refractive index (RI) stratified media~\cite{roy2014manifestations}, or by asymmetric scattering~\cite{Mondal15,PhysRevLett.130.243802,wu2025shape}. However, these approaches have limited tunability, especially in controlling the direction of rotation of a single particle. 

Here, we demonstrate that a tightly focused linearly polarized Gaussian beam can induce tunable spinning motion of trapped microparticles -- both clockwise (CW) and counterclockwise (CCW) -- provided the illumination geometry lacks axial symmetry. Specifically, we show that spinning motion arises when (1) the sample chamber is tilted (Fig. \ref{fig:schematic} (b)), (2) the trapping beam is laterally displaced (Fig. \ref{fig:schematic} (c)), or (3) the beam has an elliptical cross-section (Fig. \ref{fig:schematic} (c)). These asymmetries are among the most commonly encountered and effectively capture the essential features of realistic experimental conditions in optical tweezers. These effects originate from SOI of light under tight focusing, which dictates the redistribution of SAM within the focal volume. Without these asymmetries, the opposite spin contributions due to spin-Hall effect in the transverse (XY) focal plane exactly cancel, resulting in zero net spin and, consequently, no particle rotation (Fig. \ref{fig:schematic}(d)).  Nonetheless, as mentioned earlier -- spin-dependent responses can be deliberately induced through an engineered sample chamber, such as using multiple layers with high RI gradients that spatially align regions of high optical intensity (needed for trapping) with localized spin-angular-momentum densities (needed for rotation) \cite{roy2014manifestations}, or by employing multiple beams to selectively suppress specific spin contributions~\cite{stilgoe2022controlled}. In contrast, asymmetric illumination breaks the balance between right- and left-handed helicity (spin) densities. This imbalance manifests as an asymmetric spin Hall effect (Fig. \ref{fig:schematic} (e)) in the transverse focal plane, producing a finite positive or negative net spin density over the spatial region occupied by the trapped particle.


\section{Theoretical framework}

To model the focal field distributions arising from the aforementioned asymmetric illumination conditions, we employ the vectorial diffraction framework originally developed by Debye and Wolf \cite{richards1959electromagnetic,wolf1959electromagnetic,torok1997electromagnetic,munro2018tool}. This formalism depicts the focal field as a superposition of numerous plane-wave components emanating from the reference sphere which represents the aplanatic lens (here objective lens). The polarization evolution in the high–numerical-aperture (NA) system is governed by the lens transfer function, which accounts for Fresnel amplitude reflection and transmission at successive interfaces. This approach thus provides a realistic description of practical focusing systems involving multiple media separated by planar interfaces along the optical axis. The time-independent electric field at a point $p$ $(\mathbf{r}_{p})$ in a homogeneous medium of RI $n_0$ is:
\begin{equation}
   \mathbf{E}(\mathbf{r}_p) = -i \beta \int \int_{\Omega} \frac{\boldsymbol{\epsilon}(k_x, k_y)}{k_z} e^{in_0 k_0 (\mathbf{k} \cdot \mathbf{r}_p)} \, dk_x \, dk_y.
   \label{eq:placeholder}
\end{equation}
Here, $\beta = \frac{n_0 k_0 f}{2\pi}$, $k_0$ denotes the free-space wave number, $n_0$ the refractive index of the medium in the focal region. The term $\boldsymbol{\epsilon}(k_x, k_y)$ represents the electric field spectrum defined on the reference Gaussian sphere, while $\mathbf{k} = (k_x, k_y, k_z)=\frac{\boldsymbol{k}}{|\boldsymbol{k}|}$ specifies the unit propagation direction of an individual geometric ray (or partial ray). The integration extends over the solid angle $\Omega$ defined by the numerical aperture (NA) of the objective lens. For conciseness, the present discussion is restricted to the transformation of the electric field components only. The temporal factor is assumed to have the form $\exp{[-i\omega t]}$, where $\omega$ is the frequency of the monochromatic field. 


\begin{figure}
\centering\includegraphics[width=1 \linewidth]{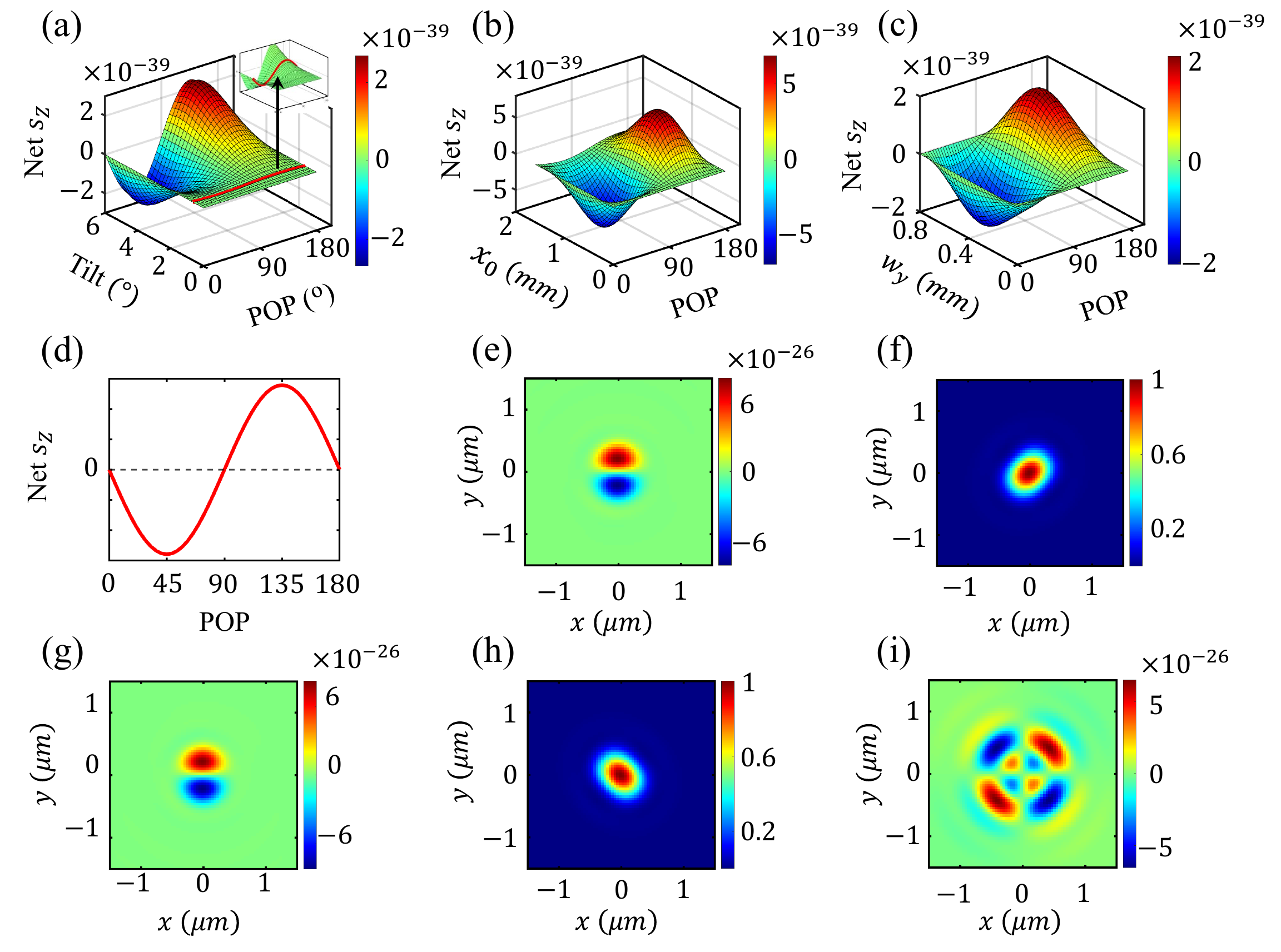}
\caption{Surface plots showing the variation of the net spin longitudinal component $S_z$ with the angle of the plane of polarization for three types of asymmetries: (a) tilting angle of the sample chamber, (b) transverse displacement of the trapping beam, and (c) beam radius along $y$ (i.e. $w_y$). The red line plots in (a) and in the inset of (a) represent the simulated variation of  $S_z$ for the experimentally determined tilt angle of $0.92^{\circ}$. (d) The variation of $S_z$ with the angle of the plane of polarization (in the transverse XY plane) exhibits the same characteristic behavior for all asymmetries. Distributions of (e) the spin component $s_z$ (net $S_z$ is positive-valued) and (f) the corresponding normalized intensity profile for a  $45^{\circ}$ plane of polarization, and similarly (g) $s_z$ (net $S_z$ is negative-valued) and (h) the normalized intensity profile for a $135^{\circ}$ plane of polarization. The same tilt angle of $0.92^{\circ}$ is used. The $s_z$ distributions in (e) and (g) exhibit two regions of opposite helicity, in contrast to the symmetric $s_z$ distribution observed for a perfectly horizontal chamber ($\theta_{\text{tilt}}=0$), which shows the usual spin Hall pattern in the focal plane of optical tweezers.}
\label{fig:simulation1}
\end{figure}
The focusing problem in a tilted stratified medium is decomposed into two independent processes: focusing is described by the Debye–Wolf formalism, while propagation across the stratified layers is treated using the transfer-matrix formalism, as first demonstrated by Munro et al. \cite{munro2018tool}. By construction, the Debye–Wolf integral yields the electric field in the laboratory frame $L_{1}$ (say). For a tilted sample chamber \cite{sauvikpra,sauvikapl,munro2018tool}, it is more convenient to evaluate the field in a rotated frame $L_{2}$ (say), whose axes are rotated by an angle $\theta_{\text{tilt}}$. Accordingly, all relevant vector quantities —- the incident electric field $\bf{E}$, the propagation vectors $\mathbf{k}$, and the particle position vector $\mathbf{r}_{p}(x,y,z)$ -- are transformed using a $3\times3$ rotation matrix $R$ that incorporates both the rotation axis and the tilt angle $\theta_{\text{tilt}}$. The field is computed in the rotated frame and subsequently transformed back to the laboratory frame via $\textbf{E}_{final}=R^{-1}\textbf{E}^{'}=R^{-1}\textbf{E}R$. This similarity-transformation approach enables efficient modeling of tilted stratified media through coordinate rotation. Specifically, the tilt is implemented as a rotation about the $y$-axis, although, in general, it may be oriented along any arbitrary direction. A detailed exposition of the complete methodology is provided in Refs. \cite{sauvikpra,sauvikapl,munro2018tool} and is not repeated here.


The incident Gaussian beam, at the back focal plane of the focusing lens, is characterized by:
\begin{equation}
\mathbf{E}_{in}=Ae^{-\left[\frac{(x-x_0)^2}{w_x^2} + \frac{(y-y_0)^2}{w_y^2}\right]}
\end{equation}
Where $A$ is the amplitude, $w_x$ and $w_y$ are the beam widths (radii) along the $x$ and $y$-axes, respectively, and the beam is centered at point $(x_0,y_0)$ on the back focal plane. It is worth noting that, in addition to sample chamber tilting, the present method is well equipped to handle beam displacement and elliptical beam profiles. The magnitude of the beam displacement and the beam-radius ratio $w_x:w_y$ for elliptically shaped incident beams can be precisely controlled through the aforementioned equation. The connections and underlying similarities among these three approaches are elucidated in the following discussion. Having found the electric field in the focal region, the local spin density is calculated using the relation:
\begin{equation}
    \mathbf{s}=\frac{1}{4\omega} \text{Im}[\varepsilon_{0}n^2 \mathbf{E}^{*}\times \mathbf{E} + \mu_{0} \mathbf{H}^{*}\times \mathbf{H}]
\end{equation}
Here, $\varepsilon_{0}$ and $\mu_{0}$ are the permittivity and permeability of free space. Notably, only the longitudinal component of the spin density $s_{z}$ is used to analyze the rotation dynamics of the motion.

\section{Results \& Discussions}

\subsection{Simulation results}


The above-mentioned method is implemented in a MATLAB program and two sets of numerical simulations are performed. In the first set, each asymmetry was introduced individually to examine its isolated effect.  The second set considered pairwise combinations of asymmetries to analyze their coupled influence. It is to be noted that there are four layers in the sample chamber: objective immersion oil ($\text{RI} = 1.516$), coverslip ($\text{RI} = 1.516$), aqueous solution of the sample ($\text{RI}_{\text{water}} = 1.33$) and the top glass-slide ($\text{RI} = 1.516$) (Fig. \ref{fig:schematic}(a)). All the  simulations are performed at the same focal plane (deep inside the aqueous or the third layer; $25\mu\text{m}$ thick) obtained for the ideal case ($z_{focus}=-1.08\mu \text{m}$) —no tilt, no displacement, and a perfect Gaussian beam with equal beam waists along the $x$ and $y$ directions ($w_x = w_y$). The axial positions of the oil–coverslip, cover slip–aqueous sample, and aqueous sample–top glass interfaces are defined as $z_1 =-165 \mu\text{m}$, $z_2 = -5 \mu\text{m}$, and $z_3 = 20\mu\text{m}$, respectively. The analysis proceeds by evaluating the integral of the longitudinal spin component $\text{s}_z$ over the region occupied by the particle i.e., net ${s}_{z}=\iint \text{s}_{z}(x,y) \,dx\,dy$. For simulation purposes, this region is approximated as a $4\times4 \mu \text{m}^2 $ area in the focal plane, since the dominant contribution arises predominantly from the localized zone near the actual focus. It is worth noting that, in ensuring coherency, the parameters -- namely the interface positions ($z_1, z_2, z_3$), the refractive indices (RI) of the layers, the tilt angle $\theta_{tilt} (=0.92^{\circ})$, the beam waist $w_{0} (=0.8mm)$, and the wavelength $\lambda (=1064nm)$ -- are adopted directly from the experimental values presented in the following section.

For a tilted sample chamber, the surface distribution of the net $s_z$ across the trapping plane is analyzed by varying both the tilt angle and the plane of polarization (POP) of the incident linearly polarized beam. It is observed that, in the absence of tilt, the net $s_z$ remains zero throughout the rotation of the polarization plane (Figs. \ref{fig:simulation1}(a), (i)). However, for any finite tilt angle, the net $s_z$ exhibits a sinusoidal dependence, attaining extrema at polarization angles of $45^{\circ}$ and $135^{\circ}$. At $45^{\circ}$, $s_z$ assumes a negative value, indicating a dominant negative-helicity contribution, whereas at $135^{\circ}$ it becomes positive, corresponding to a dominant positive-helicity component in the focal plane (Fig. \ref{fig:simulation1}(a)). The occurrence of extrema at $45^{\circ}$ and $135^{\circ}$ implies that, over a full $360^{\circ}$ rotation of the plane of polarization (or equivalently a $180^{\circ}$ rotation of a half-wave plate controlling the plane of polarization), the net $s_z$ flips between positive and negative four times.



The spatial distributions of the longitudinal spin angular momentum ($s_z$) and the focal intensity are examined for a specific tilt angle. In particular, a tilt of $0.92^\circ$ is chosen, corresponding to the configuration employed in the experimental validation of our simulations (discussed later). It is observed that, although the spatial profiles of $s_z$ appear identical for polarization angles of $45^{\circ}$ (Fig.~\ref{fig:simulation1}(e)) and $135^{\circ}$ (Fig.~\ref{fig:simulation1}(g)), their overall effects are entirely opposite. Specifically, the negative integrated value in the $45^{\circ}$ case implies a counter-clockwise (CCW) rotational motion, whereas the positive integrated value in the $135^{\circ}$ case should lead to a clockwise (CW) rotation of the trapped microparticles — thereby resulting in opposite rotational dynamics. The corresponding intensity profiles for $45^{\circ}$ and $135^{\circ}$ are shown in Figs. \ref{fig:simulation1}(f), (h),  respectively. The elongation of the focal spot along the $45^{\circ}$ and $135^{\circ}$ directions for the respective polarization states is clearly visible. Furthermore, owing to the small tilt angle, the intensity profiles exhibit minimal deformation. It is to be noted that, in later simulations, we do vary the the tilting angle to look for maximal effects, but do not increase the value beyond  $6^{\circ}$. This is because at larger tilt angles, the focal point may shift inside the cover slip, rendering normal optical tweezers operation impractical.


\begin{figure}
\centering\includegraphics[width=1 \linewidth]{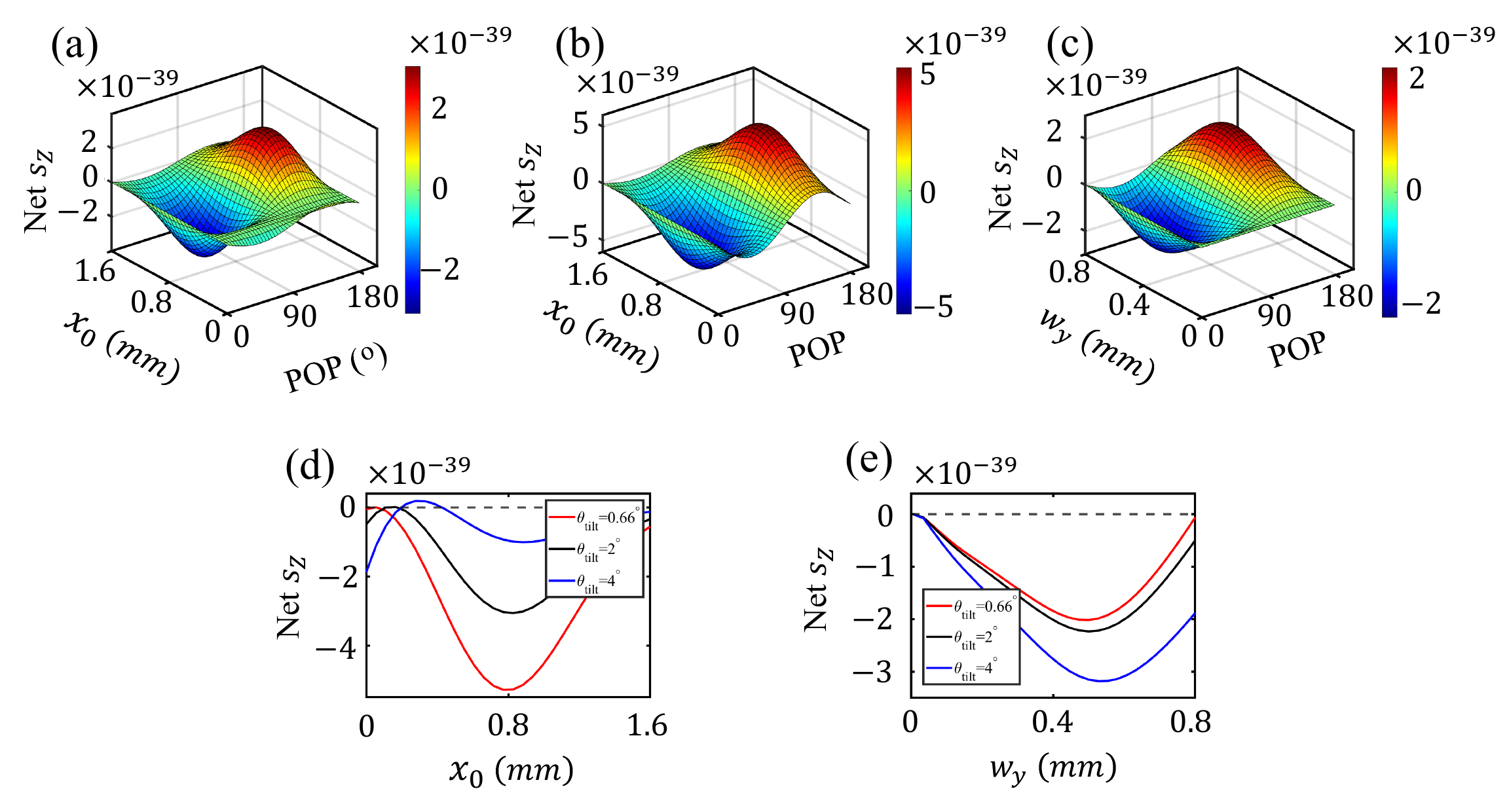}
\caption{(a) Surface plot showing the combined effect of tilt and lateral displacement on the net $s_z$, evaluated as a function of the plane-polarization angle and beam displacement along $x$. The tilt angle is fixed at $2^{\circ}$. (b) Surface plot illustrating the combined effect of beam displacement for an elliptical Gaussian beam on the net $s_z$, shown as a function of the plane-polarization angle and displacement along $y$. The beam widths are $w_x = w_0 = 0.8\,\mathrm{mm}$ and $w_y = w_x/2 = 0.4\,\mathrm{mm}$. (c) Surface plot depicting the combined influence of tilt and beam ellipticity on the net $s_z$, plotted against the plane-polarization angle and the minor-axis radius $w_y$. The tilt is fixed at $2^{\circ}$ and $w_x = w_0 = 0.8\,\mathrm{mm}$. (d) Variation of the net $s_z$ with beam displacement for three different tilt angles $\theta_{\mathrm{tilt}}$. (e) Variation of the net $s_z$ with the minor-axis beam radius $w_y$ for three different tilt angles $\theta_{\mathrm{tilt}}$.}
\label{fig:simulation2}
\end{figure}

Next, we analyze the influence of beam displacement on the net $s_z$. When there is no displacement and the beam axis coincides with the optical axis of the lens, the net $s_z$ is zero and remains insensitive to the polarization plane of the incident beam (Fig. \ref{fig:simulation1}(i)). This configuration has been extensively studied and is characterized by a symmetric, diagonally opposite spatial separation of the positive and negative helicity components-a phenomenon commonly referred to as the optical Hall effect. However, upon introducing a lateral displacement of the beam, the net $s_z$ exhibits a sinusoidal variation, as shown in Fig. \ref{fig:simulation1}(b). In this case, $s_z$ is negative for polarization angles between $0^{\circ}$ and $90^{\circ}$, and positive between $90^{\circ}$ and $180^{\circ}$, attaining negative and positive extrema at $45^{\circ}$ and $135^{\circ}$, respectively. For a fixed displacement, this dependence of $s_z$ on the polarization angle closely mirrors the behavior observed for a fixed tilt angle. The extremum observed near $0.8\text{mm}$ can be directly attributed to the experimental geometry: a beam with waist $w_0=0.8\text{mm}$ is displaced by $0.8\text{mm}$, matching the radius of the back aperture. As a result, nearly half of the beam is physically blocked at the aperture, while the remaining portion is transmitted and sampled, giving rise to the observed asymmetric response at the extrema. Notably, displacing the beam from a chosen origin introduces extrinsic orbital angular momentum \cite{AllenExternalOAMPRL2002}; however, the present study focuses on the spin distribution near the focal plane, as described by the vectorial Debye–Wolf formalism.

For the elliptical beam, the beam waist along the $x$-axis is fixed at $w_x = 0.8\text{mm}$, while the waist along the $y$-axis, $w_y$, is varied (Fig. \ref{fig:simulation1}(c)). When $w_x = w_y$, the situation corresponds to the ideal symmetric case—analogous to the initial configurations in the previous two scenarios (tilt and displacement) — yielding a net $s_z$ of zero for all polarization angles. As $w_y$ deviates from $w_x$, the net $s_z$ exhibits a sinusoidal dependence on the polarization plane. In this case, $s_z$ is negative for polarization angles between $0^{\circ}$ and $90^{\circ}$, and positive between $90^{\circ}$ and $180^{\circ}$, with extrema occurring at $45^{\circ}$ and $135^{\circ}$. For a fixed ratio of $w_x$ to $w_y$ (i.e., a fixed elliptical-ness), this dependence closely resembles the behavior observed for the other two asymmetries.

Among the three asymmetries discussed above, two -- namely the beam displacement and elliptical beam shape -- arise from asymmetric distributions in the field amplitudes of the incident $\mathbf{k}$-vectors (or partial waves). In both situations, the amplitude distribution departs from that of an ideal Gaussian beam perfectly aligned with the optical axis and incident on a sample chamber held strictly horizontal. For a displaced beam, the central maximum shifts laterally, enhancing certain off-axis $\mathbf{k}$-vectors with larger convergence angles $\theta$, thereby causing the focusing lens to sample the Gaussian profile asymmetrically. In the case of an elliptic beam, the field amplitudes of the horizontal plane waves differ from those of the vertical plane waves.

For a tilted sample chamber, however, the amplitudes of the partial waves emerging from the lens remain unchanged; instead, the plane waves deviate from their initial planes of incidence due to the inclined interfaces within the chamber. These waves then overlap to produce the resultant field distribution experienced by the trapped particle. Despite procedural differences among the modalities, axial symmetry is broken in all three cases, and the net longitudinal spin ($s_z$) in the focal plane exhibits similar behavior (Fig. \ref{fig:simulation1}(d)). Consequently, the third Stokes parameter $(S_3)$—typically used to characterize the spin Hall shift—no longer retains its usual functional form
\begin{equation}
S_3 = A(\rho)\sin(2\phi)
\end{equation}
which exhibits two maxima and two minima due to the $\sin(2\phi)$ dependence. This leads to a deviation of the spin distribution from the conventional spin Hall separation observed in optical tweezers. It is worth emphasizing that, in all asymmetric illumination modalities, the conservation of angular momentum is not violated; rather, it is redistributed either within the focal plane or, in the case of beam displacement, transferred to the rigid lens mount.


Another interesting aspect is the combination of these asymmetric modalities. In Fig. \ref{fig:simulation1}, we observe opposite rotational as well as no-rotation dynamics, demonstrating the tunability of the net spin angular momentum arising independently from sample tilt, beam displacement, and beam ellipticity. However, this tunability can be further enhanced when these asymmetries are combined. Figure \ref{fig:simulation2}(a) shows that, for a fixed sample-chamber tilt angle ($\theta_{\text{tilt}}=2^{\circ}$), the net $s_z$ can be continuously tuned, and the positions of its extrema can be predictably shifted by varying the beam displacement. This displacement-induced tunability is illustrated in Fig. \ref{fig:simulation2}(d) for three different tilt angles using a Gaussian beam polarized at $45^{\circ}$. Similar controls can be achieved using a displaced elliptical beam ($w_x=w_0=0.8,\text{mm}$, $w_y=w_0/2=0.4,\text{mm}$), as shown in Fig. \ref{fig:simulation2}(b), or by combining beam ellipticity with a tilted sample chamber, as shown in Fig. \ref{fig:simulation2}(c) for $\theta_{\text{tilt}}=2^{\circ}$. Moreover, the net $s_z$ increases with increasing tilt angle for an elliptical beam, as evidenced in Fig. \ref{fig:simulation2}(e) for $\theta_{\text{tilt}}=0.66^{\circ}$, $2^{\circ}$, and $4^{\circ}$.

An important feature of the tilted sample chamber is that, as the tilt angle increases, the maximum values of the net positive and negative $s_z$ also increase. In other words, the difference between the maximum values of the generated left circular and right circular polarization (LCP and RCP) consistently grows with tilt. This behavior contrasts with displacement- and elliptical beam–induced asymmetry, where the effect typically diminishes beyond a certain point. The monotonic growth of net $s_z$ with tilt offers a practical advantage: experiments performed at essentially any nonzero tilt angle should, in principle, be able to induce rotation in microparticles and thereby validate the numerical predictions for asymmetric illumination. In the following section, we present the corresponding experimental results.

\subsection{Experimental results}
\begin{figure}
    \centering
    \includegraphics[width=0.98\linewidth]{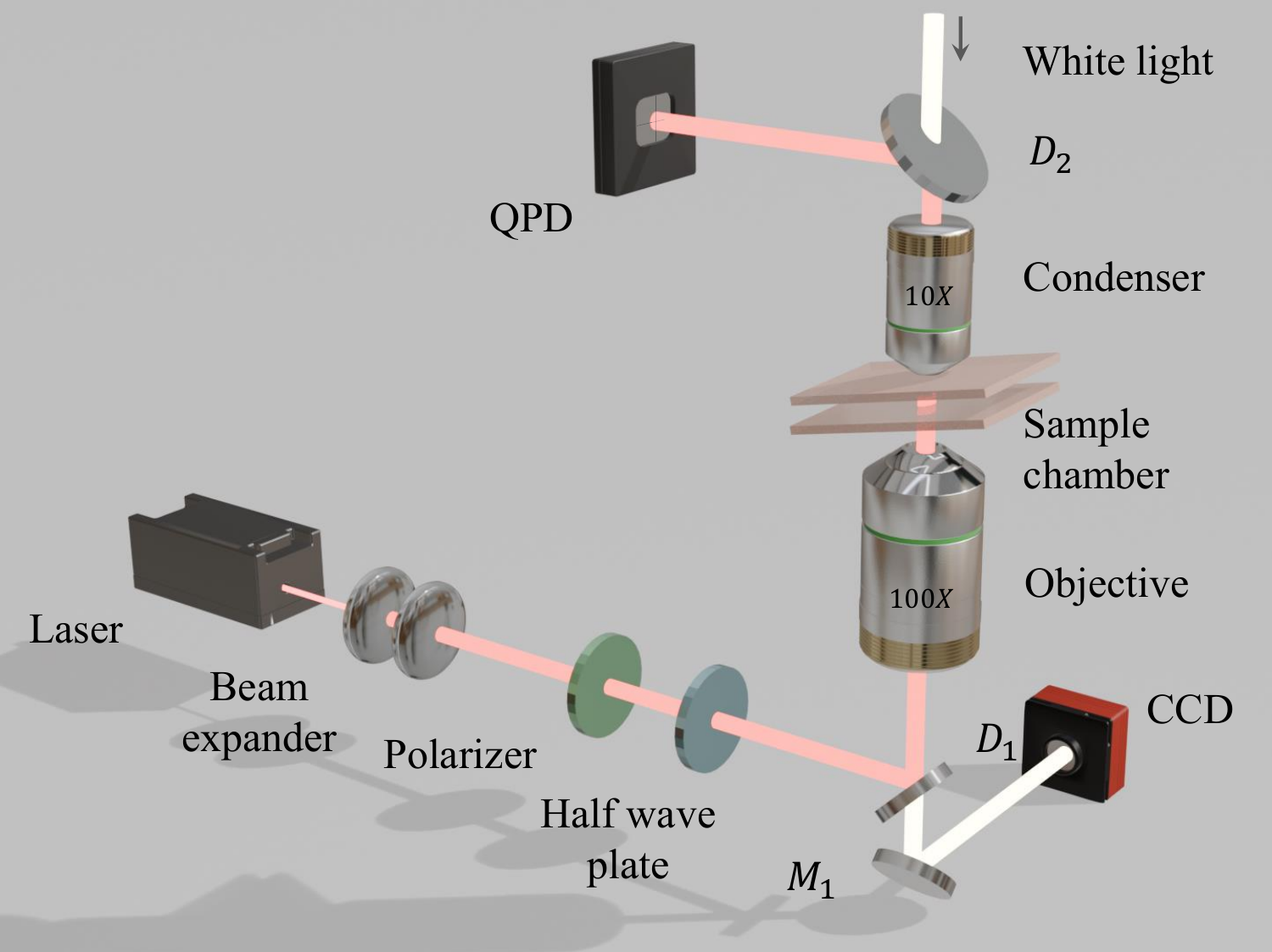}
    \caption{Experimental setup of optical tweezers with tilted sample chamber. $M$: mirror, $D:$ dichroic mirror.}
    \label{fig:exptsetup}
\end{figure}

In our experimental setup, a linearly polarized Gaussian beam (wavelength, $\lambda = 1064,\mathrm{nm}, w_0=0.8mm$) is tightly focused into a stratified sample chamber as shown in Fig. \ref{fig:exptsetup}. Nematic 5CB liquid crystal (LC) particles dispersed in deionized water are used as the probe (or trapped) particle. A 1 wt$\%$ polyvinyl alcohol (PVA) solution is also added to promote the formation of bipolar LC particles \cite{AFernandezPRL2004}. 5CB liquid crystals exhibit relatively high birefringence $(n_o\approx 1.515, n_e\approx 1.667, \Delta n=n_e-n_o\approx 0.152 \text{ at } 1064\mathrm{nm})$ \cite{RI5CB}, making them particularly suitable for facilitating efficient angular momentum transfer from the optical field to the particle. The plane of polarization (POP) of the incident beam is rotated in discrete steps over a full $360^{\circ}$ using a half-wave plate (HWP) mounted on a motorized rotation stage and positioned between the laser source and the objective lens. For a perfectly horizontal (zero-tilt) sample chamber, the particle follows the changing polarization direction, settling into a new stable orientation at each step \cite{brasselet2009}, without the onset of any continuous rotation. Next, the sample chamber is intentionally tilted by $\theta_{\text{tilt}}=0.92^{\circ}$ since this is the maximum angle we can achieve experimentally, keeping in mind the fact that for high-numerical-aperture objective lenses, the limited working distance at the exit pupil restricts the use of larger tilting angles; otherwise, the focal point shifts inside the cover slip, thereby compromising the trapping functionality of the optical tweezers.

\begin{figure}
\centering\includegraphics[width=1 \linewidth]{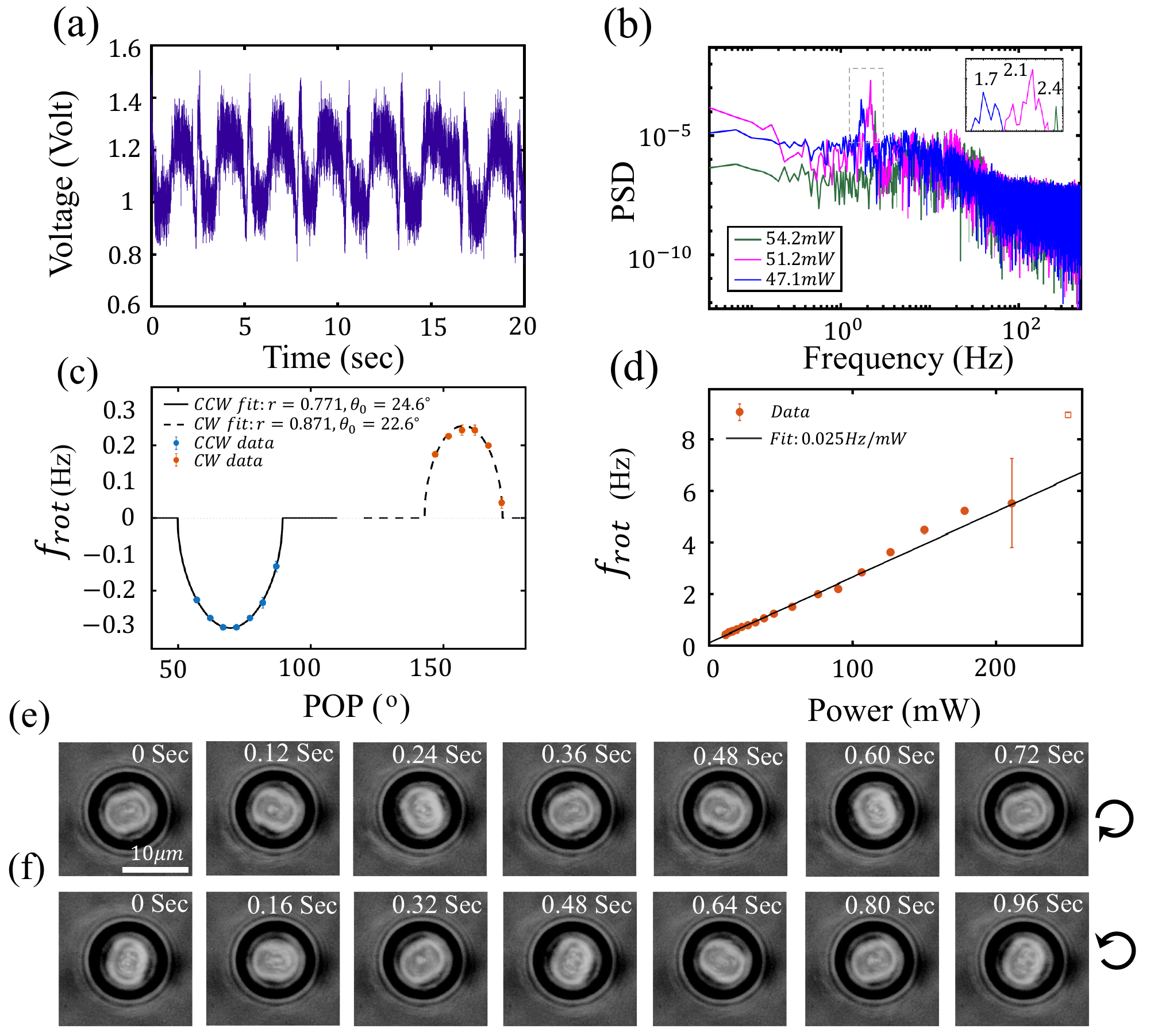}
\caption{(a) Time trace of the voltage signal recorded from one quadrant of the photodiode. (b) Power spectral densities measured at three different incident powers. Inset: Corresponding rotational frequencies of the trapped particle. (c) Rotation frequency as a function of the angle of the plane of polarization (POP); negative values indicate counterclockwise (CCW) rotation. The lines show separate fits of Eq. (\ref{eq:frequency}) to the CCW and clockwise (CW) rotation windows, with $(r,\theta_0)=(0.771,24.6^{\circ})$ and $(0.871,22.6^{\circ})$, respectively. (d) Rotation frequency as a function of the incident power at the sample, varied using a variable neutral-density (ND) filter, measured at the CCW optimum. The line represents a linear fit; the open square denotes the highest-power data point, which was excluded from the fit. Time-lapse images showing (e) CW and (f) CCW rotation of LC particles trapped at the focus, for incident polarization angles of $135^{\circ}$ and $45^{\circ}$, respectively, with $\theta_{\text{tilt}}=0.92^{\circ}$.}
\label{fig:experiment}
\end{figure}
The scattered field from the trapped LC particle is detected using a quadrant photodiode (QPD) placed in the forward-scattering configuration. A representative time series data of the voltage signal from a single quadrant of the QPD, exhibiting periodic motion, is shown in Fig. \ref{fig:experiment}(a). The signal oscillates on a broadband background due to translational fluctuations in the trap and detector noise. The rotational frequencies of the particle ($f_{rot}$) as a function of the angle of rotation of the HWP are then extracted from the power spectral density (PSD) of the voltage difference between the principal or counter-diagonal quadrants of the QPD (see Figs. \ref{fig:experiment}(b), (c)). For a trapped non-rotating spherical particle, the power spectral density (PSD) follows a Lorentzian profile \cite{berg2004power,neuman2004optical}. Upon the onset of rotation, distinct peaks emerge in the PSD at the rotational frequency, enabling the particle's rotation frequency ($f_{rot}$) to be determined experimentally. Figure \ref{fig:experiment}(b) presents the PSDs from the same particle for three different incident laser powers. The PSD peak frequencies $(f_{peak})$ are 1.7 Hz, 2.1 Hz, and 2.4 Hz for incident powers of 47.1 mW, 51.2 mW, and 54.2 mW , respectively (inset of Fig.  \ref{fig:experiment}(b)). The bipolar director configuration that 5CB adopts in water \cite{brasselet2009,mosallaeipour2010lJMO} is head–tail symmetric, so the signal is invariant under a rotation by $\pi$ and repeats twice per turn. The rotation frequency is therefore half the PSD peak, $f_{rot} = f_{peak}/2$ .

When the HWP is aligned at $0^{\circ}$ with respect to the incoming linearly polarized beam, no rotational motion of the trapped particle is observed. As the HWP is rotated, the first onset of counterclockwise (CCW) rotation occurs at $\theta_{\text{HWP}} = 17.5^{\circ}$, corresponding to a polarization rotation of $\theta_{\text{pol}} = 2\times\theta_{\text{HWP}} = 35^{\circ}$ (Fig. \ref{fig:experiment} (c)). The CCW rotation persists up to $\theta_{\text{pol}} = 64^{\circ}$, as evident from Fig. \ref{fig:experiment}(c).   Within this interval, the rotational frequency increases gradually from $\theta_{\text{pol}} = 34^{\circ}$, reaches a maximum at $\theta_{\text{pol}}^{\text{max}} = 44^{\circ}$, and subsequently decreases until $\theta_{\text{pol}} = 64^{\circ}$ (Fig. \ref{fig:experiment} (c)). A time lapse sequence of CCW rotational motions are shown in Fig. \ref{fig:experiment}(e).

Upon further rotation of the HWP, clockwise (CW) spinning of the LC particle is observed (see Supplementary Video 1) in the polarization range $124^{\circ} \le \theta_{\text{pol}} \le 149^{\circ}$, as shown in Figs. \ref{fig:experiment}(c) and (f), with the peak CW rotational frequency occurring at $\theta_{\text{pol}}^{\text{max}} = 134^{\circ}$. Thus, we obtain excellent agreement between the simulated polarization angles corresponding to maximal opposite spin directions and the experimentally observed conditions for stationary and rotating states (see Fig. \ref{fig:simulation1}(d) and Fig. \ref{fig:experiment}(c)). Moreover, over a complete $360^{\circ}$ rotation of the HWP, we observe eight alternating transitions between CCW and CW motion, as we clearly demonstrated in our simulations (see Fig.~\ref{fig:simulation1}(d)). Note that only the $0^{\circ}$ to $180^{\circ}$ polarization range is shown in Fig \ref{fig:simulation1}(d). This periodic reversal arises because a rotation of the HWP by an angle $\theta_{\text{HWP}}$ induces a $2\theta_{\text{HWP}}$ rotation of the polarization plane of the incident beam. Consequently, the maxima in rotational frequency are expected to occur at intervals of $2\times 360/8 = 90^{\circ}$, which can also be seen in Fig. \ref{fig:experiment}(c). The optical power is deliberately kept low ($\sim 11.2$ mW) to visually track the reversal in rotation direction.

Although the simulated and experimental results show good agreement in the overall spin behavior, the experimentally measured rotational frequency deviates noticeably from an ideal sinusoidal dependence. This deviation is evident from the absence of rotation within the range $65^{\circ} \leq \theta_{\text{pol}} \leq 123^{\circ}$, the narrower rotation windows compared with those predicted by an ideal sine curve, and the unequal widths and heights of the CW and CCW windows. The dead zones (absence of rotation) reflect the existence of a second optical torque: a birefringent particle in a linearly polarized field tends to turn towards the polarization plane by an alignment torque $\tau_\text{align} = -\tau_a\sin\left[2(\alpha-\theta_{\mathrm{pol}})\right]$~\cite{rotationdraghalina}, where $\alpha$ is the in-plane orientation of the optic axis and $\tau_a \propto P\Delta n$, $P$ being the incident power and $\Delta n$ is the birefringence mentioned earlier \cite{HalinaNature1998,la2004PRL}. The alignment torque has a period-\(\pi\) due to the head–tail symmetry of the particle's optical axis. To describe the rotational dynamics of the birefringent LC particle under the combined influence of the spin torque $\tau_s$, optical alignment torque $\tau_\text{align}$, and thermal fluctuations arising from Brownian motion in the surrounding liquid bath, we employ an overdamped Langevin equation:
\begin{equation}
\gamma_r \dot{\alpha}
=
\tau_s(\theta_{\mathrm{pol}})
-
\tau_a \sin\left[2(\alpha-\theta_{\mathrm{pol}})\right]
+
\sqrt{2\gamma_r k_B T}\,\xi(t),
\label{eq:Langevin}
\end{equation}
where  $\gamma_r$ is the rotational friction coefficient, $k_B$ is the Boltzmann constant, $T$ is the temperature of the surrounding medium, and $\xi(t)$ is Gaussian white noise with zero mean and unit strength, $\langle \xi(t)\,\xi(t') \rangle = \delta(t-t')$. The noise amplitude $\sqrt{2\gamma_r k_B T}$ is set by the fluctuation--dissipation theorem. The spin torque follows the net spin density,
\begin{equation}
\tau_s(\theta_{\mathrm{pol}})
=
-\tau_{s,\max}
\sin\left[2(\theta_{\mathrm{pol}}-\theta_0)\right],
\label{eq:spintorque}
\end{equation}
where $\theta_0$ is an experimental offset of the polarization zero, namely the angle between the half-wave-plate reading and the plane of polarization of the incident beam. The two deterministic torques in Eq.~(\ref{eq:Langevin}) can be written as the gradient of a single effective potential, $\gamma_r\dot{\alpha} = -\partial U/\partial\alpha + \sqrt{2\gamma_r k_B T}\,\xi(t)$, so that the orientation of the particle performs overdamped Brownian motion in a tilted (biased) periodic potential~\cite{reimann2001PRL,pedaci2011NaturePhysics},
\begin{equation}
U(\alpha)
=
-\tau_s\alpha
-
\frac{\tau_a}{2}
\cos\left[2(\alpha-\theta_{\mathrm{pol}})\right].
\label{eq:6}
\end{equation}
For $\tau_a>|\tau_s|$, $U(\alpha)$ has minima and the particle is locked at
$
\sin\left[2(\alpha^*-\theta_{\mathrm{pol}})\right]
=
\tau_s/\tau_a
$. Thus, any change in the plane of polarization will be accompanied by a mere reorientation of the trapped particle. This defines the dead zone. On the contrary, when $|\tau_s|>\tau_a$ \cite{HalinaNature1998,pedaci2011NaturePhysics}, a continuous rotation will be observed. Integrating Eq.~(\ref{eq:Langevin}) without noise over one period of the alignment torque gives the mean rotation frequency  $f_{\mathrm{rot}} = \sqrt{\tau_s^2-\tau_a^2}/(2\pi\gamma_r)$. 
Thermal fluctuations, included in Eq.~(\ref{eq:Langevin}), are negligible for the large droplets studied here, except in a narrow range of $\theta_{\mathrm{pol}}$ around the onset of rotation, $|\tau_s| = \tau_a$, where the barrier of $U(\alpha)$ vanishes and thermally activated phase slips smear the transition~\cite{reimann2001PRL}. Since this range is small compared with the angular step of our measurements, we analyse the deterministic mean rotation rate.
With negative frequency denoting CCW rotation, the explicit form of $f_{rot}(\theta_{pol})$ is:
\begin{equation}
f_{\mathrm{rot}}(\theta_{\mathrm{pol}})
=
-f_0\,
\operatorname{sgn}
\left[
\sin 2(\theta_{\mathrm{pol}}-\theta_0)
\right]
\sqrt{
\sin^2 2(\theta_{\mathrm{pol}}-\theta_0)-r^2
}
\label{eq:frequency}
\end{equation}
where, $r=\tau_a/\tau_{s,\max}$ and
$f_0=\tau_{s,\max}/(2\pi\gamma_r)$.
Thus, each rotation window has width $90^\circ-\sin^{-1}{r}$, and the peak rate is $\sqrt{1-r^2}\,f_0$. 
Separate fits of Eq.~(\ref{eq:frequency}) to the CCW and CW windows (solid and dashed lines in Fig. \ref{fig:experiment}(c), respectively) yield the following parameters (see Table \ref{tab:fig4c_fitting}).
\begin{table}[htbp]
\centering
\caption{Fitting parameters for the CCW and CW rotation windows in Fig. \ref{fig:experiment}(c).}
\begin{tabular}{cccccc}
\hline
Window &
$r=\tau_a/\tau_{s,\max}$ &
$\theta_0$ &
$f_0$ (Hz) &
Width &
Peak $|f_{\mathrm{rot}}|$ (Hz) \\
\hline
CCW &
$0.771\pm0.003$ &
$24.6^\circ\pm0.1^\circ$ &
$0.474\pm0.004$ &
$39.6^\circ$ &
$0.30$ \\
CW &
$0.871\pm0.006$ &
$22.6^\circ\pm0.3^\circ$ &
$0.52\pm0.02$ & 
$29.4^\circ$ &
$0.25$ \\
\hline
\end{tabular}
\label{tab:fig4c_fitting}
\end{table}
The alignment torque amounts to $77-87\%$ of the maximum spin torque $(\tau_{s,max})$, substantially narrowing the rotation windows. Consequently, even at the optimal polarization angle, the rotation rate reaches only $\sqrt{1-r^2}=64\%$ (CCW) and $49\%$ (CW) of the corresponding rates in the absence of alignment torque. Near threshold, the particle slows on each corrugation of $U(\alpha)$, which accounts for the non-sinusoidal waveform in Fig. \ref{fig:experiment}(a). Thus, the rotation threshold is set by the competition between the two optical torques, with the alignment torque nearly matching the spin torque. Sustained rotation occurs only when the tilt-induced spin torque overcomes the alignment torque; otherwise, the droplet remains orientationally locked. Viscous drag determines the rotation rate but does not produce the dead zones. Also, the CW window is narrower and slower than the CCW window, indicating an approximately $11\%$ weaker CW spin torque relative to the alignment torque, consistent with the slower CW rotation observed in Supplementary Video 1. To understand the weaker CW spin torque, a small polarization-independent spin torque $\tau_b$ acting on the LC particle is added: $
\tau_s
=
-\tau_{s,\max}
\sin\left[2(\theta_{\mathrm{pol}}-\theta_0)\right]
+\tau_b.
$
Then, the ratio
$
\dfrac{r_{\mathrm{CW}}}{r_{\mathrm{CCW}}}
=
\dfrac{1+b}{1-b}
$
yields
$b=\tau_b /\tau_{s,\max}\approx 0.06$. This small polarization-independent contribution to the spin torque ($\sim 6\%$), likely arising from residual beam ellipticity, manifests as an asymmetry in the rotation windows. The values of the offset $\theta_{0}$ are approximately $2^{\circ}$ apart for the two windows.

Throughout the experiment, since only the HWP is rotated, the optical power at the trapping plane remains nearly constant. The maximum deviation observed for different plane of polarization is $ \sim 4.6\%$ at the trapping plane. Note that smaller particles exhibit a lower onset angle of the polarization plane and higher rotational frequencies. Furthermore, a linear dependence of the rotational frequency ($f_{rot}=\kappa P +c,$ $c=0.03\text{Hz}$) on the incident optical power is observed, as shown in Fig. \ref{fig:experiment}(d) for a fixed angle of polarization, $\theta_{\text{pol}} = 45^{\circ}$. Since both torques increase proportionally with the incident power $P$, their ratio $r$ remains unchanged, thus $f_0$ increases linearly with $P$, as evident from Eq. (\ref{eq:frequency}). 

Converting the measured rotation rates into absolute torques requires accurate calibration of both the droplet’s rotational friction and the optical power at the particle, each of which carries systematic uncertainty. The power measured after the objective, with the power meter placed in air, underestimates the power at the particle because it does not capture large-angle rays (NA $>1$). 
The rotational friction scales with the cube of the droplet diameter, and the Stokes value $8\pi\eta R^3$, with $\eta$ the viscosity of water and $R$ the droplet radius, holds only for a rigid sphere with a no-slip surface; a liquid-crystal droplet may deviate from it\cite{saito2022ScientificReports,saito2024SoftMatter}. 
In addition, converting the photodiode signal into a rotation frequency relies on the assumed symmetry of the bipolar director configuration. Given these uncertainties, absolute torque values are not reported. Instead, we focus on quantities independent of these calibrations: the alignment-to-spin torque ratio inferred from the widths of the rotation windows in Fig. \ref{fig:experiment}(c), the rotation threshold, and the linear scaling of rotation frequency with optical power in Fig. \ref{fig:experiment}(d).
\begin{figure}
    \centering
    \includegraphics[width=1\linewidth]{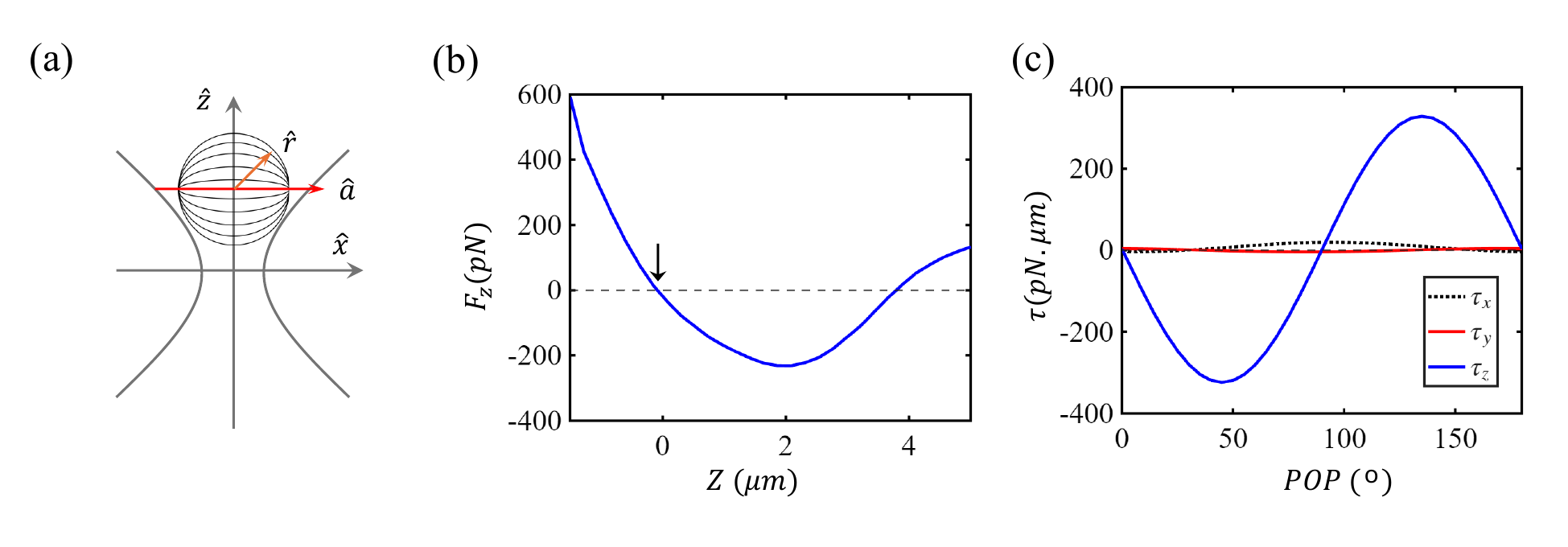}
    \caption{(a) Schematic of the stable orientation of a bipolar liquid-crystal (LC) particle in the optical trap, with \(\hat{\mathbf{a}}\) denoting its bipolar symmetry axis. (b) Axial force \(F_z\) on a spherical LC particle of radius \(3~\mu\mathrm{m}\) as a function of its axial position \(Z\). The black arrow indicates the stable trapping position at \(Z=-0.155~\mu\mathrm{m}\), where \(\partial F_z/\partial Z<0\). (c) Three torque components, \(\tau_x\), \(\tau_y\), and \(\tau_z\), acting on the particle at the stable trapping position as functions of the plane-of-polarization angle (POP). The optical power injected in the simulations is \(1~\mathrm{Watt}\).}
    \label{fig:forcetorque}
\end{figure}
It is worth noting that the Maxwell stress tensor (MST) calculation yields the total torque acting on the particle, whereas uniquely separating it into alignment and spin contributions is challenging. A 3D FDTD simulation (Ansys Lumerical FDTD) is performed, considering the particle geometry, anisotropic material properties, and interaction with the tightly focused Debye–Wolf electromagnetic field. Owing to computational constraints, we considered a spherical LC particle of radius \(3\,\mu\mathrm{m}\). The optical anisotropy of the LC is modelled using a spatially varying uniaxial dielectric tensor. In the local director frame, it is given by $\boldsymbol{\varepsilon}_{\mathrm{local}}
= \varepsilon_0\,\operatorname{diag}
\left[n_o^2,n_o^2,n_e^2\right]$ \cite{RI5CB}. In the laboratory frame, the dielectric tensor takes the form
$
\boldsymbol{\varepsilon}(\mathbf r)
=\varepsilon_0\left[
n_o^2\mathbf I+(n_e^2-n_o^2)
\mathbf n(\mathbf r)\mathbf n^{T}(\mathbf r)
\right],
$
where \(\mathbf n(\mathbf r)\) is the local bipolar director. We adopt an idealized bipolar director configuration in which the director is tangential to the spherical surface and follows the bipolar symmetry axis \(\mathbf a\). It is defined as
$
\mathbf n(\mathbf r)=
\dfrac{\mathbf a-(\mathbf a\cdot\hat{\mathbf r})\hat{\mathbf r}}
{\left|\mathbf a-(\mathbf a\cdot\hat{\mathbf r})\hat{\mathbf r}\right|},
$
with \(\hat{\mathbf r}=\mathbf r/|\mathbf r|\). The simulated total torque as functions of the plane of polarization for a bipolar particle held in its stable orientation (Fig. \ref{fig:forcetorque}(a)) \cite{brasselet2009,mosallaeipour2010lJMO} and trapping position is shown in Fig. \ref{fig:forcetorque}. The stable axial trapping position  is determined from the MST-based force calculation using the stability condition $\partial F_z/\partial z < 0$ (black arrow at $z=-0.155\mu m$ in Fig. \ref{fig:forcetorque}(b)). The longitudinal torque, $\tau_z$, which drives rotation about the $z$-axis, exhibits a dependence on the polarization angle consistent with that predicted by the net longitudinal spin angular momentum of the optical field (Figs. \ref{fig:forcetorque}(c), \ref{fig:experiment}(c), \ref{fig:simulation1}(d)). In contrast, the transverse components, $\tau_x$ and $\tau_y$, are much smaller than $\tau_z$. Overall, the study demonstrates an experimentally viable strategy for inducing both clockwise and counterclockwise rotation and for continuously tuning the rotational frequency at nearly identical trapping power.

\section{Conclusion}
In summary, we demonstrate that a single tightly focused, linearly polarized Gaussian beam -- despite carrying zero net intrinsic spin angular momentum -- can exert a controllable optical torque on birefringent microparticles when axial symmetry in the illumination geometry is deliberately broken. Through a combination of experiments and vectorial diffraction–based simulations, we show that commonly encountered asymmetries such as sample-chamber tilt, lateral beam displacement, and beam ellipticity all lead to an imbalance in the spin Hall–type separation of left- and right-handed helicity components in the focal region. This imbalance gives rise to a finite net longitudinal spin density, whose magnitude and sign can be continuously tuned by rotating the incident plane of polarization, thereby enabling reversible clockwise and counterclockwise particle rotation at nearly identical optical power. For a birefringent probe, the resulting spin torque competes with the alignment torque
exerted by the linearly polarized field. Rotation therefore occurs only in windows of polarization angle where the former exceeds the latter, and the polarization angle switches the particle between rotating in either sense and being orientationally locked. Importantly, the apparently disparate mechanisms are unified by a single physical principle: the breaking of axial symmetry in a high–numerical-aperture focusing system, resulting in an asymmetric spin-Hall effect. Our results establish an experimentally accessible and minimal route to optical rotation and controllable torque generation without invoking beams that intrinsically carry angular momentum, such as circularly, radially or azimuthally polarized beams -- thereby broadening the conceptual framework of spin–orbit interactions in tightly focused fields and opening new possibilities for robust, flexible, and power-efficient optomechanical control at the micro- and nanoscale. 

\section{DATA AVAILABILITY}
The data and code supporting the findings of this study are available from the authors upon reasonable request.

\section{Supporting Information}
See the supplementary material for some of the experimental aspects of the work in further detail. Video 1: This video shows clockwise, stationary and counterclockwise motion of a trapped birefringent particle in response to a linearly polarized Gaussian beam. The plane of polarization is rotated by a half wave plate in discrete steps. 

\section{Acknowledgments}
SR is thankful to the Department of Science and Technology (DST), Government of India for INSPIRE fellowship. BD is thankful to the Ministry Of Education of Government of India for financial support through the Prime Minister’s Research Fellowship (PMRF) grant.


\bibliographystyle{apsrev4-2}
\bibliography{apsamp}

\end{document}